\documentclass[oneside,10pt]{article}

\usepackage{graphicx}
\usepackage{changepage}
\usepackage{authblk}
\usepackage{listings}

\newenvironment{myenv1}{\begin{adjustwidth}{0.0cm}{}}{\end{adjustwidth}}
\newenvironment{myenv2}{\begin{adjustwidth}{0.5cm}{}}{\end{adjustwidth}}
\newenvironment{myenv3}{\begin{adjustwidth}{1.0cm}{}}{\end{adjustwidth}}
\newenvironment{myenv4}{\begin{adjustwidth}{1.5cm}{}}{\end{adjustwidth}}

\usepackage{xpatch}
\xpatchcmd{\author}{\relax#1\relax}{\relax\detokenize{#1}\relax}{}{}

\begin{document}
\title{{\LARGE Visual C++ Implementation of Sinogram-based Adaptive Iterative Reconstruction for Sparse View X-Ray CT}}
\author[\empty]{D. Trinca\textsuperscript{a,}\thanks{Corresponding author, email: dntrinca@yahoo.com}}
\author[b]{Y. Zhong}
\author[b]{Y. Wang}
\author[b]{T. Mamyrbayev}
\author[c]{E. Libin}
\affil[a]{Sc Piretus Prod Srl, Osoi, jud. Iasi, Romania}
\affil[b]{Russian-Chinese Laboratory of Radiation Control and Inspection, Tomsk Polytechnic University, Russian Federation}
\affil[c]{Research Institute of Applied Mathematics and Mechanics, Tomsk State University, Russian Federation}
\date{}
\maketitle

\begin{abstract}
With the availability of more powerful computing processors, iterative reconstruction algorithms have recently been successfully implemented as an approach to achieving significant dose reduction in X-ray CT. In this report, we descrive our recent work on developing an adaptive iterative reconstruction algorithm for X-ray CT, that is shown to provide results comparable to those obtained by proprietary algorithms, both in terms of reconstruction accuracy and execution time. The described algorithm is thus provided for free to the scientific community, for regular use, and for possible further optimization.
\end{abstract}

\section{Introduction}
Iterative reconstruction algorithms \cite{Beister-2012} for X-ray computed tomography (CT) \cite{Natterer-2001} have been extensively applied and studied recently as an approach for lowering radiation exposure to X-rays \cite{Beister-2012} during clinical examinations. Iterative techniques have been used for a long time in nuclear medicine, but only during the last few years several manufacturers have made available and suggested the use of iterative methods for routine CT imaging that simultaneously provide for acceptable image quality with detectability of low-contrast objects and significant dose reduction.

Comparing to the filtered back-projection algorithm (FBP) \cite{Natterer-2001}, iterative reconstruction algorithms such as ART \cite{ART1}, SART \cite{SART1}, and SIRT \cite{SIRT1} normally take significantly more time for obtaining reconstructed tomograms of comparable accuracy. However, recently proposed algorithms such as the Adaptive Statistical Iterative Reconstruction (ASIR) \cite{Silva-2010,Xu-2013} and the Sinogram-Affirmed Iterative Reconstruction (SAFIRE)  \cite{Baker-2012,Wang-2012,Schulz-2013,Pontana-2013,Wang-2013,Becce-2013,Shin-2013,Pontana-2015,Siemens1} provide clinically acceptable results within a reconstruction time comparable with that of the FBP algorithm. Statistical reconstruction algorithms such as ASIR have been criticized for their "plastic-like" reconstruction \cite{Siemens1}. In the majority of case studies undertaken, the potential dose reduction of SAFIRE is around $50\%$, and in some cases around $60-65\%$ (for example, for a number of chest CT examinations). For some other case studies, the potential dose reduction that can be achieved is around $35-40\%$, as in

(1) \cite{Becce-2013}, where the authors have applied computed tomography of cervical spine, and compared reduced-dose SAFIRE with standard dose of $100\%$ FBP; it has been concluded that the dose can be reduced up to $40\%$ with SAFIRE in order to provide results comparable with $100\%$ dose FBP, although "the former protocol provides lower image quality of the soft tissues and vertebrae" which means that the dose reduction would be actually slightly lower than 40\%;

(2) \cite{Shin-2013}, where comparisons between SAFIRE and automated kV modulation (CARE kV) for abdominal CT imaging are made, and it is stated that "dose can be decreased up to 41.3\%";

(3) \cite{Schulz-2013}, where performance of iterative image reconstruction of the paranasal sinuses is studied, and it is stated that "Subjective quality evaluation of the noise-adapted images showed preference for those acquired at 100\% tube current with FBP (4.7-5.0) versus 50\% dose with SAFIRE (3.4-4.4)".

Adaptive iterative methods such as those used in the ASIR and SAFIRE algorithms could be as well applied not only in clinical CT, but also in other routine X-ray examinations such as non-destructive testing of materials, for speed-up. As the algorithmic details of these methods are proprietary, we aim in this report to describe our proposal for an adaptive iterative reconstruction algorithm, that is shown to have potential dose reduction of $50\%$ with reconstruction times comparable to the SAFIRE algorithm. After presenting in detail the algorithm and providing illustrative results, we then discuss a number of possible optimizations.
\section{Adaptive Iterative Reconstruction}
In this section we describe, step-by-step, the details of our proposed adaptive iterative reconstruction algorithm. Let $\mu_{1}$ be the density matrix (of size {\it nx} lines by {\it ny} columns) to be reconstructed from sinogram $S$, ${\it nd}$ the number of detectors, and ${\it np}$ the number of projection angles. $S[i,j]$ is the sinogram value corresponding to detector $i$, for the $j$-th projection angle. There are two main steps in the reconstruction process:
\begin{enumerate}
\item initialization of the reconstruction matrix $\mu_{1}$ with initial solution,
\item and iterations.
\end{enumerate}

The detailed description of the reconstruction is as follows. Before the two main steps proceed, we have the following initializations of variables:
\begin{displaymath}
{\it \mu_{1}[i,j]\leftarrow{0.0}\textnormal{ , }1\leq{i}\leq{\it nx}, 1\leq{j}\leq{\it ny}},
\end{displaymath}

\begin{displaymath}
{\it \mu_{2}[i,j]\leftarrow{0.0}\textnormal{ , }1\leq{i}\leq{\it nx}, 1\leq{j}\leq{\it ny}},
\end{displaymath}

\begin{displaymath}
{\it o_{1}[i,j]\leftarrow{0.0}\textnormal{ , }1\leq{i}\leq{\it nx}, 1\leq{j}\leq{\it ny}}.
\end{displaymath}
After these initializations of variables, the initialization of the reconstruction matrix $\mu_{1}$ with initial solution is done as in the algorithm given in Fig. \ref{fig:initialization}, shown in pseudocode.
\begin{figure}[!t]
\begin{myenv1}
{\it for every $j$, $1\leq{j}\leq{\it np}$ do}
\end{myenv1}
\begin{myenv2}
{\it for every $i$, $1\leq{i}\leq{\it nd}$ do}
\end{myenv2}
\begin{myenv3}
{\it let $B_{i,j}$ be the X-ray beam that corresponds to the sinogram value $S[i,j]$;}
\end{myenv3}
\vspace{0.1cm}
\begin{myenv3}
{\it let $L_{i,j}$, $C_{i,j}$ be 2 vectors with $\mu_{1}[L_{i,j}[1],C_{i,j}[1]]$, \ldots, $\mu_{1}[L_{i,j}[c_{i,j}],C_{i,j}[c_{i,j}]]$ being all $c_{i,j}$ entries of $\mu_{1}$ that correspond to the beam $B_{i,j}$;}
\end{myenv3}
\vspace{0.1cm}
\begin{myenv3}
{\it let ${\it Seg}_{i,j}$ be a vector such that ${\it Seg}_{i,j}[k]$ is the length of the segment corresponding to the path that $B_{i,j}$ follows through the entry $\mu_{1}[L_{i,j}[k],C_{i,j}[k]]$, $1\leq{k}\leq{c_{i,j}}$;}
\end{myenv3}
\vspace{0.1cm}
\begin{myenv3}
{\it ${\it SUM}\leftarrow{0.0}$;}
\end{myenv3}
\vspace{0.1cm}
\begin{myenv3}
{\it for every $k$, $1\leq{k}\leq{c_{i,j}}$ do}
\end{myenv3}
\begin{myenv4}
{\it ${\it SUM}\leftarrow{\it SUM}+{\it Seg}_{i,j}[k]$;}
\end{myenv4}
\begin{myenv3}
{\it endfor}
\end{myenv3}
\vspace{0.1cm}
\begin{myenv3}
{\it for every $k$, $1\leq{k}\leq{c_{i,j}}$ do}
\end{myenv3}
\begin{myenv4}
{\it $o_{1}[L_{i,j}[k],C_{i,j}[k]]\leftarrow{o_{1}[L_{i,j}[k],C_{i,j}[k]]}+{\it Seg}_{i,j}[k]$;}
\end{myenv4}
\begin{myenv3}
{\it endfor}
\end{myenv3}
\vspace{0.1cm}
\begin{myenv3}
{\it for every $k$, $1\leq{k}\leq{c_{i,j}}$ do}
\end{myenv3}
\begin{myenv4}
{\it $\mu_{1}[L_{i,j}[k],C_{i,j}[k]]\leftarrow{\mu_{1}[L_{i,j}[k],C_{i,j}[k]]}+{\it Seg}_{i,j}[k]*(S[i,j]/{\it SUM})$;}
\end{myenv4}
\begin{myenv3}
{\it endfor}
\end{myenv3}
\begin{myenv2}
{\it endfor}
\end{myenv2}
\begin{myenv1}
{\it endfor}
\end{myenv1}
\begin{myenv1}
{\it for every $i$, $1\leq{i}\leq{\it nx}$ do}
\end{myenv1}
\begin{myenv2}
{\it for every $j$, $1\leq{j}\leq{\it ny}$ do}
\end{myenv2}
\begin{myenv3}
{\it $\mu_{1}[i,j]\leftarrow\mu_{1}[i,j]/o_{1}[i,j]$;}
\end{myenv3}
\begin{myenv2}
{\it endfor}
\end{myenv2}
\begin{myenv1}
{\it endfor}
\end{myenv1}
\caption{Pseudocode of initialization}
\label{fig:initialization}
\end{figure}

The initialization with the initial solution is very simple, and can be explained as follows.

For every $j$ and $i$, $1\leq{j}\leq{\it np}$, $1\leq{i}\leq{\it nd}$: let $B_{i,j}$ be the X-ray beam that corresponds to the sinogram value $S[i,j]$, and suppose that $B_{i,j}$ goes through $c_{i,j}$ entries of $\mu_{1}$. Let $L_{i,j}$, $C_{i,j}$ be two vectors such that
\begin{displaymath}
\mu_{1}[L_{i,j}[1],C_{i,j}[1]], \ldots, \mu_{1}[L_{i,j}[c_{i,j}],C_{i,j}[c_{i,j}]]
\end{displaymath}
are all entries of $\mu_{1}$ that correspond to the beam $B_{i,j}$. For the current $j$ and $i$, ${\it SUM}$ is the sum of all segments corresponding to beam $B_{i,j}$ (the entries of $\mu_{1}$ that correspond to the current beam $B_{i,j}$, and the associated segments, are computed in the same way when the sinogram was formed). Then, for every $k$, $1\leq{k}\leq{c_{i,j}}$, $o_{1}[L_{i,j}[k],C_{i,j}[k]]$ is the sum of all segments (that is, from all beams) corresponding to the entry $\mu_{1}[L_{i,j}[k],C_{i,j}[k]]$; so, for the current beam $B_{i,j}$ we add the current segments ${\it Seg}_{i,j}[1]$, $\dots$, ${\it Seg}_{i,j}[c_{i,j}]$ to the corresponding entries in $o_{1}$ (that is, to $o_{1}[L_{i,j}[1],C_{i,j}[1]]$, $\ldots$, $o_{1}[L_{i,j}[c_{i,j}],C_{i,j}[c_{i,j}]]$). Since ${\it SUM}$ is the sum of all segments corresponding to the current beam $B_{i,j}$, it follows that
\begin{displaymath}
S[i,j]/{\it SUM}
\end{displaymath}
is the current average sinogram value per unit of segment. But, the current beam $B_{i,j}$ goes through $\mu_{1}[L_{i,j}[k],C_{i,j}[k]]$ for distance ${\it Seg}_{i,j}[k]$, and not for unit distance 1.0, so it follows that
\begin{displaymath}
{\it Seg}_{i,j}[k]*(S[i,j]/{\it SUM})
\end{displaymath}
is an approximation of the contribution of $\mu_{1}[L_{i,j}[k],C_{i,j}[k]]$ to $S[i,j]$. We add to $\mu_{1}[L_{i,j}[k],C_{i,j}[k]]$ all these approximations.

For every $i$ and $j$, $1\leq{i}\leq{\it nx}$, $1\leq{j}\leq{\it ny}$ we finally initialize $\mu_{1}[i,j]$ by dividing the sum of approximations of contributions by the sum of segments corresponding to $\mu_{1}[i,j]$.

After the initialization step follows the iterations step; at each iteration the algorithm from Fig. \ref{fig:iteration}, shown in pseudocode, is executed.
\begin{figure}[t]
\begin{myenv1}
{\it for every $j$, $1\leq{j}\leq{\it np}$ do}
\end{myenv1}
\begin{myenv2}
{\it for every $i$, $1\leq{i}\leq{\it nd}$ do}
\end{myenv2}
\begin{myenv3}
{\it let $B_{i,j}$ be the X-ray beam that corresponds to the sinogram value $S[i,j]$;}
\end{myenv3}
\vspace{0.1cm}
\begin{myenv3}
{\it let $L_{i,j}$, $C_{i,j}$ be 2 vectors with $\mu_{1}[L_{i,j}[1],C_{i,j}[1]]$, \ldots, $\mu_{1}[L_{i,j}[c_{i,j}],C_{i,j}[c_{i,j}]]$ being all $c_{i,j}$ entries of $\mu_{1}$ that correspond to the beam $B_{i,j}$;}
\end{myenv3}
\vspace{0.1cm}
\begin{myenv3}
{\it let ${\it Seg}_{i,j}$ be a vector such that ${\it Seg}_{i,j}[k]$ is the length of the segment corresponding to the path that $B_{i,j}$ follows through the entry $\mu_{1}[L_{i,j}[k],C_{i,j}[k]]$, $1\leq{k}\leq{c_{i,j}}$;}
\end{myenv3}
\vspace{0.1cm}
\begin{myenv3}
{\it ${\it Sit}\leftarrow{0.0}$;}
\end{myenv3}
\vspace{0.1cm}
\begin{myenv3}
{\it for every $k$, $1\leq{k}\leq{c_{i,j}}$ do}
\end{myenv3}
\begin{myenv4}
{\it ${\it Sit}\leftarrow{\it Sit}+{\it Seg}_{i,j}[k]*\mu_{1}[L_{i,j}[k],C_{i,j}[k]]$;}
\end{myenv4}
\begin{myenv3}
{\it endfor}
\end{myenv3}
\vspace{0.1cm}
\begin{myenv3}
{\it for every $k$, $1\leq{k}\leq{c_{i,j}}$ do}
\end{myenv3}
\begin{myenv4}
{\it $\mu_{2}[L_{i,j}[k],C_{i,j}[k]]\leftarrow{\mu_{2}[L_{i,j}[k],C_{i,j}[k]]}+{\it Seg}_{i,j}[k]*\mu_{1}[L_{i,j}[k],C_{i,j}[k]]*(S[i,j]/{\it Sit})$;}
\end{myenv4}
\begin{myenv3}
{\it endfor}
\end{myenv3}
\begin{myenv2}
{\it endfor}
\end{myenv2}
\begin{myenv1}
{\it endfor}
\end{myenv1}
\begin{myenv1}
{\it for every $i$, $1\leq{i}\leq{\it nx}$ do}
\end{myenv1}
\begin{myenv2}
{\it for every $j$, $1\leq{j}\leq{\it ny}$ do}
\end{myenv2}
\begin{myenv3}
{\it $\mu_{1}[i,j]\leftarrow\mu_{2}[i,j]/o_{1}[i,j]$;}
\end{myenv3}
\vspace{0.1cm}
\begin{myenv3}
{\it $\mu_{2}[i,j]\leftarrow{0.0}$;}
\end{myenv3}
\begin{myenv2}
{\it endfor}
\end{myenv2}
\begin{myenv1}
{\it endfor}
\end{myenv1}
\caption{Pseudocode of the algorithm executed at each iteration}
\label{fig:iteration}
\end{figure}

The code executed at each iteration is similar with the initialization with initial solution, but with some difference.

For every $j$ and $i$, $1\leq{j}\leq{\it np}$, $1\leq{i}\leq{\it nd}$: $L_{i,j}$, $C_{i,j}$, ${\it Seg}_{i,j}$ are already calculated from the initialization with initial solution; then, the corresponding situation ${\it Sit}$ in the reconstruction, for the current X-ray beam $B_{i,j}$, is calculated;
\begin{displaymath}
\mu_{1}[L_{i,j}[k],C_{i,j}[k]]*(S[i,j]/Sit)
\end{displaymath}
is the corrected value of $\mu_{1}[L_{i,j}[k],C_{i,j}[k]]$ for the current detector and projection angle, and
\begin{displaymath}
{\it Seg}_{i,j}[k]*\mu_{1}[L_{i,j}[k],C_{i,j}[k]]*(S[i,j]/{\it Sit})
\end{displaymath}
is the exact contribution of this corrected value of $\mu_{1}[L_{i,j}[k],C_{i,j}[k]]$; in the variable $\mu_{2}$, we add these exact contributions of the corrected values.

For every $i$ and $j$, $1\leq{i}\leq{\it nx}$, $1\leq{j}\leq{\it ny}$ we finally compute $\mu_{1}[i,j]$ by dividing the current sum of exact contributions of the corrected values by the sum of segments corresponding to $\mu_{1}[i,j]$. Also, $\mu_{2}$ is re-initialized for the next iteration.

The iterations stop when the difference between reconstructions from two consecutive iterations reaches a predefined threshold.
\section{Results Obtained by the Proposed Adaptive Iterative Reconstruction}
\begin{figure}[t]
\centering
\includegraphics{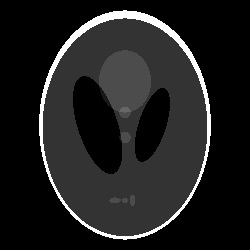}
\caption{Cross-section of size 250 by 250 pixels}
\label{fig:1}
\end{figure}

\begin{figure}[t]
\centering
\includegraphics{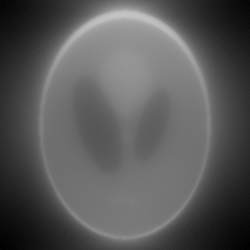}
\caption{Result of initialization with initial solution (198 projection angles)}
\label{fig:Result-of-initialsolution}
\end{figure}

\begin{figure}[t]
\centering
\includegraphics{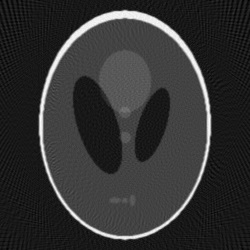}
\caption{Result of reconstruction by the FBP algorithm (198 projection angles)}
\label{fig:fbp198}
\end{figure}

In this section, we examine the performance of the described adaptive iterative reconstruction algorithm. Consider the Shepp-Logan tomogram shown in Fig. \ref{fig:1}, of size 250 by 250 pixels. This tomogram has been generated using the MATLAB software \cite{MATLAB1}, using the command
\begin{displaymath}
\textnormal{phantom('Modified Shepp-Logan', 250)}.
\end{displaymath}
For this cross-section, consider the following parameters:
\begin{enumerate}
\item distance from fan-beam source to origin of rotation of inspected object = 800,
\item distance from fan-beam source to line of detectors = 1500,
\item number of detectors equally spaced on the detector line = 359,
\item number of projection angles = 198.
\end{enumerate}
The corresponding sinogram is a matrix with 359 lines and 198 columns. For this set of parameters, and fan-beam scanning with detectors arranged equally spaced on the detector line, the initial solution is as given in Fig. \ref{fig:Result-of-initialsolution}; the result obtained by the the FBP algorithm is shown in Fig. \ref{fig:fbp198}; as it is visible, there are many artifacts.

\begin{figure}[t]
\centerline{
\includegraphics{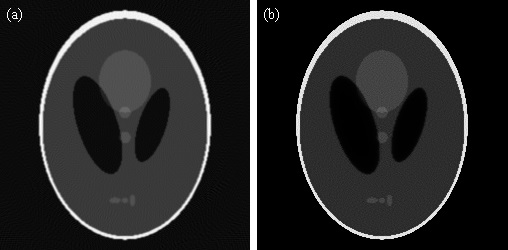}
}
\caption{(a) result of reconstruction by the FBP algorithm, 360 projection angles; (b) result of reconstruction by the adaptive iterative reconstruction algorithm, 198 projection angles}
\label{fig:Comparison1}
\end{figure}

The result by the adaptive iterative reconstruction, for the same set of parameters, and run with 285 iterations, is shown in Fig. \ref{fig:Comparison1} (b); in Fig. \ref{fig:Comparison1} (a), the result obtained by the FBP algorithm is shown, but for 360 projection angles. The result obtained with the adaptive iterative reconstruction (Fig. \ref{fig:Comparison1} (b), 198 projection angles) is of the same quality as the result obtained by the FBP algorithm (Fig. \ref{fig:Comparison1} (a), 360 projection angles), which means an X-ray dose reduction of $45\%$. The result for the adaptive iterative reconstruction has been obtained using a desktop computer with Intel Xeon E5-2697 v2, 2.70GHz processor, by parallelization on all 12 cores available, using Visual Studio 2013 software. The execution time for the inspected tomogram was about 1.75 seconds. From all the tests which have been run, it has been concluded that the dose can be reduced up to $50\%$ for the considered Shepp-Logan tomogram, but using $30-40\%$ more iterations.

This result is thus comparable with those obtained by the SAFIRE algorithm, both in terms of dose reduction and execution time (the SAFIRE algorithm takes, to the best of our knowledge, around 1 second for tomograms of this size), but as opposed to the SAFIRE algorithm, whose details are proprietary, the described adaptive iterative reconstruction algorithm is provided for free to the community of researchers working on tomographic reconstruction (not only in the medical sector where the SAFIRE algorithm is used, but also for non-destructive testing of objects, etc.), for regular use.

The execution time of the described  iterative reconstruction algorithm could be reduced by using more powerful processors, or on parallelization on Graphical Processing Units (GPUs).
\section{Optimizations}
In this section, we discuss possible optimizations. The execution time that we have reported in the previous section was obtained by parallelization of the proposed algorithm on multi-core processors. Both the initialization step and the iterations step can be parallelized easily by splitting the work done for all detector -- projection angle pairs equally among all cores available. For the example that we have analyzed, where we have $359$ detectors and $198$ projection angles, and if we have 11 cores available, then the first core would deal with all detector -- projection angles pairs $(i,j)$ for $1\leq{i}\leq{359}$, $1\leq{j}\leq{18}$, the second core would deal with all detector -- projection angles pairs $(i,j)$ for $1\leq{i}\leq{359}$, $19\leq{j}\leq{36}$, etc. For the iterations step, clearly the calculation of $\mu_{1}$ from the variables $\mu_{2}$, $o_{1}$ at the end of each iteration would be done only by one of the cores, and the others would wait before all cores start the next iteration, as each core needs the same matrix $\mu_{1}$ at the start of each iteration.

Besides parallelization on multi-core processors, one other possible optimization in practice could be the following: in the code executed at each iteration, for each of the ${\it nd}*{\it np}$ X-ray beams, a number of entries of the matrix $\mu_{1}$ are used for calculating ${\it Sit}$, the situation for the current X-ray beam. If some of these entries are exactly $0.0$ then the addition of the respective
\begin{displaymath}
{\it Seg}_{i,j}[k]*\mu_{1}[L_{i,j}[k],C_{i,j}[k]]
\end{displaymath}
terms to the ${\it Sit}$ variable becomes useless. Also, after calculating the ${\it Sit}$ variable, the addition of the respective
\begin{displaymath}
{\it Seg}_{i,j}[k]*\mu_{1}[L_{i,j}[k],C_{i,j}[k]]*(S[i,j]/{\it Sit})
\end{displaymath}
terms to the $\mu_{2}$ variable also becomes useless. Therefore, it is desirable to not examine the respective $0.0$ entries of $\mu_{1}$. This could be realized by running, once at every few iterations, a test that checks which entries of $\mu_{1}$ have become $0.0$ and eliminate them from the $L_{i,j}$, $C_{i,j}$ vectors, for all ${\it nd}*{\it np}$ detector -- projection angle pairs. Eliminating these $0.0$ entries is a correct procedure, as once an entry has become $0.0$, it will remain $0.0$ until the end of all iterations, regardless of how many iterations are run. However, for the Shepp-Logan tomogram of size 250 by 250 that we have tested, none of the entries has reached the exact value $0.0$, but for other tomograms it may happen.

Another possible optimization would be to use a different initial solution for the described adaptive iterative reconstruction algorithm.
\section{Conclusions}
The second generation of iterative reconstruction algorithms for X-ray CT, such as the ASIR and SAFIRE adaptive algorithms, have been followed up by many case studies where it is shown that potential dose reduction of around 50\% can be applied as compared to the filtered back-projection algorithm. In this report, we have described our proposal for an adaptive iterative reconstruction algorithm that is shown to produce very good accuracy, is fast, and provided for free to the scientific community for regular use, and possible further improvement.
\appendix
\section{Visual C++ Implementation}
\begin{figure}[t]
\centerline{
\includegraphics{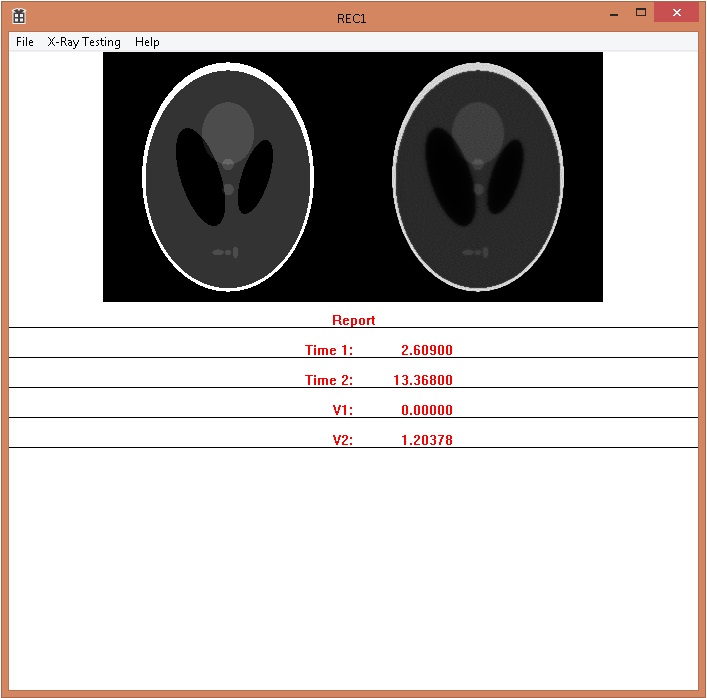}
}
\caption{Visual C++ 2013 implementation of Sinogram-based Adaptive Iterative Reconstruction: on the left hand-side is the original cross-section, and on the right hand-side is the reconstruction after 150 iterations}
\label{fig:REC1}
\end{figure}

In this section, we show the Visual C++ 2013 main code implementing the described adaptive iterative reconstruction algorithm. This is a usual Win32 application, and an example of the interface is shown in Fig. \ref{fig:REC1}. The function executed by each thread is called "MyThreadFunction". This implementation uses barriers (for synchronization of threads after each iteration during the reconstruction process), a facility that requires Windows 8 or higher.
\lstset {language=C++}
\begin{lstlisting}
#include "stdafx.h"
#include "REC1.h"
#define _USE_MATH_DEFINES
#include <math.h>
#include <time.h>


#include <vector>
#include <algorithm>


using namespace std;

#define MAX_LOADSTRING 100

// Global Variables:
HINSTANCE hInst;
TCHAR szTitle[MAX_LOADSTRING];
TCHAR szWindowClass[MAX_LOADSTRING];

const int    np = 359;
const int    nf = 180;
const int    nx = 250;
const int    ny = 250;
const double  a = 700;
const double  d = 800;
const double c1 = (2.0*M_PI) / nf;
int			 NI;

struct INF { int *Lin;int *Col;double *Seg;double SumOfSegs;int* ind;int Count; };


double *Mu1;
double *RI1;
double *RIG;
double *mO1;
INF    *Z;
double *S;


#define nThreads 4
DWORD WINAPI MyThreadFunction(LPVOID lpParam);
SYNCHRONIZATION_BARRIER barrier;



ATOM MyRegisterClass(HINSTANCE hInstance);
BOOL InitInstance(HINSTANCE, int);
LRESULT CALLBACK WndProc(HWND, UINT, WPARAM, LPARAM);
INT_PTR CALLBACK About(HWND, UINT, WPARAM, LPARAM);

int APIENTRY _tWinMain( 	_In_ HINSTANCE hInstance,
				_In_opt_ HINSTANCE hPrevInstance,
				_In_ LPTSTR lpCmdLine,
				_In_ int nCmdShow	)
{
   UNREFERENCED_PARAMETER(hPrevInstance);
   UNREFERENCED_PARAMETER(lpCmdLine);

   MSG msg;
   HACCEL hAccelTable;

   LoadString(hInstance, IDS_APP_TITLE, szTitle, MAX_LOADSTRING);
   LoadString(hInstance, IDC_REC1, szWindowClass, MAX_LOADSTRING);
   MyRegisterClass(hInstance);

   if (!InitInstance (hInstance, nCmdShow))
   {
      return FALSE;
   }

   hAccelTable = LoadAccelerators(hInstance, MAKEINTRESOURCE(IDC_REC1));

   while (GetMessage(&msg, NULL, 0, 0))
   {
      if (!TranslateAccelerator(msg.hwnd, hAccelTable, &msg))
      {
         TranslateMessage(&msg);
         DispatchMessage(&msg);
      }
   }

   return (int) msg.wParam;
}



//
//  FUNCTION: MyRegisterClass()
//
ATOM MyRegisterClass(HINSTANCE hInstance)
{
   WNDCLASSEX wcex;

   wcex.cbSize = sizeof(WNDCLASSEX);

   wcex.style		= CS_HREDRAW | CS_VREDRAW;
   wcex.lpfnWndProc	= WndProc;
   wcex.cbClsExtra	= 0;
   wcex.cbWndExtra	= 0;
   wcex.hInstance	= hInstance;
   wcex.hIcon		= LoadIcon(hInstance, MAKEINTRESOURCE(IDI_REC1));
   wcex.hCursor		= LoadCursor(NULL, IDC_ARROW);
   wcex.hbrBackground	= (HBRUSH)(COLOR_WINDOW+1);
   wcex.lpszMenuName	= MAKEINTRESOURCE(IDC_REC1);
   wcex.lpszClassName	= szWindowClass;
   wcex.hIconSm		= LoadIcon(wcex.hInstance, MAKEINTRESOURCE(IDI_SMALL));

   return RegisterClassEx(&wcex);
}

//
//   FUNCTION: InitInstance(HINSTANCE, int)
//
BOOL InitInstance(HINSTANCE hInstance, int nCmdShow)
{
   HWND hWnd;

   hInst = hInstance;

   hWnd = CreateWindow(	szWindowClass,
			szTitle,
			WS_OVERLAPPEDWINDOW,
			CW_USEDEFAULT,
			0,
			CW_USEDEFAULT,
			0,
			NULL,
			NULL,
			hInstance,
			NULL);

   if (!hWnd)
   {
      return FALSE;
   }

   ShowWindow(hWnd, SW_SHOWMAXIMIZED);
   UpdateWindow(hWnd);
   

   return TRUE;
}

//
//  FUNCTION: WndProc(HWND, UINT, WPARAM, LPARAM)
//
LRESULT CALLBACK WndProc(HWND hWnd, UINT message, WPARAM wParam, LPARAM lParam)
{
   int wmId, wmEvent;
   PAINTSTRUCT ps;
   HDC hdc;

   HDC              hdcMem;
   HGDIOBJ          hbmOld;
   BITMAPINFOHEADER bmih;
   BITMAPINFO       dbmi;
   HBITMAP          hbmp = NULL;
   BITMAP           bmp;





   static double duration1 = 0.0;
   static double duration2 = 0.0;
   static double min1       = 0.0;
   static double max1       = 0.0;
   static WCHAR Msg1[50];

   static unsigned char *cs1    = NULL;
   static unsigned char *pixels = NULL;
   void                 *bits;

   static int Status = 0;

   switch (message)
   {
      case WM_COMMAND:
         wmId = LOWORD(wParam);
         wmEvent = HIWORD(wParam);
         switch (wmId)
         {
            case ID_X_ITERATIVEMETHOD1:
            {
               SetCursor(LoadCursor(NULL, IDC_WAIT));
               InvalidateRect(hWnd, NULL, TRUE);
               Status = 0;
               MSG msg;
               msg.hwnd = hWnd;
               msg.message = WM_PAINT;
               DispatchMessage(&msg);

               clock_t start1;
               clock_t start2;

               start1 = clock();

               int i;
               int j;
               int k;
               Mu1 = (double*)malloc(nx*ny*sizeof(double));
               RI1 = (double*)malloc(nx*ny*sizeof(double));
               RIG = (double*)malloc(nx*ny*sizeof(double));
               mO1 = (double*)malloc(nx*ny*sizeof(double));
               Z   = (INF*   )malloc(np*nf*sizeof(INF   ));


               FILE *f1;


               double d1;
               if (fopen_s(&f1, "ph-250x250.txt", "r") == 0)
               {
                  for (i = 0; i < nx; i++)
                  {
                     for (j = 0; j < ny; j++)
                     {
                        fscanf_s(f1, "%lf", &d1);
                        Mu1[i*ny + j] = d1;
                     }
                  }
                  fclose(f1);
               }
               else
                  MessageBox(	hWnd,
 				(LPCWSTR)L"Problem opening the file!",
				(LPCWSTR)L"!",
				MB_OK	);
		  

               for (i = 0; i < nx; i++)
               for (j = 0; j < ny; j++)
               {
                  RI1[i*ny + j] = 0.0;
                  RIG[i*ny + j] = 0.0;
                  mO1[i*ny + j] = 0.0;
               }

               double *dc = (double*)malloc(np*sizeof(double));

               dc[0] = -((((double)np) - 1.0) / 2.0)*((a + d) / d);

               for (i = 1; i < np; i++)
                  dc[i] = dc[i - 1] + (a + d) / d;

               double x1;
               double x2;
               double y1;
               double y2;


               double *xl = (double*)malloc((ny + 1)*sizeof(double));
               double *yl = (double*)malloc((nx + 1)*sizeof(double));

               double xLoLimit, xUpLimit;
               double yLoLimit, yUpLimit;

               int nyPlusOne = ny + 1;
               int nxPlusOne = nx + 1;
               int nyMinusOne = ny - 1;
               int nxMinusOne = nx - 1;

               xLoLimit = -((((double)ny) - 1.0) / 2.0 + 0.5);
               xUpLimit = ((((double)ny) - 1.0) / 2.0 + 0.5);
               yLoLimit = -((((double)nx) - 1.0) / 2.0 + 0.5);
               yUpLimit = ((((double)nx) - 1.0) / 2.0 + 0.5);

               xl[0] = xLoLimit;
               for (i = 1; i < nyPlusOne; i++)
                  xl[i] = xl[i - 1] + 1.0;

               yl[0] = yLoLimit;
               for (i = 1; i < nxPlusOne; i++)
                  yl[i] = yl[i - 1] + 1.0;


               double xhrz;
               double yhrz;
               double xvrt;
               double yvrt;


               S = (double*)malloc(np*nf*sizeof(double));


               double f;
               double m;
               double b;
               double Sinf;
               double Sinfa;
               double Cosf;
               double Cosfa;


               double xcm;
               double ycm;
               int Lin;
               int Col;
               double seg;

               double *xV = (double*)malloc(((ny + 1) + (nx + 1))*sizeof(double));
               double *yV = (double*)malloc(((ny + 1) + (nx + 1))*sizeof(double));
               vector<pair<double, double>> V;
               int nEl;

               double Var1 = (((double)ny) - 1.0) / 2.0 + 0.5;
               double Var2 = (((double)nx) - 1.0) / 2.0 + 0.5;

               for (i = 0; i < nf; i++)
               for (j = 0; j < np; j++)
                  S[i*np + j] = 0.0;

               for (i=0;i<nf;i++)
               {
                  f=i*c1;

                  Sinf=sin(f); Sinfa=Sinf*a;
                  Cosf=cos(f); Cosfa=Cosf*a;

                  x1 = (-d)*(-Sinf);
                  y1 = (-d)*(Cosf);

                  for (j = 0; j < np; j++)
                  {
                     Z[j*nf + i].Count = 0;
                     Z[j*nf + i].Lin = NULL;
                     Z[j*nf + i].Col = NULL;
                     Z[j*nf + i].ind = NULL;
                     Z[j*nf + i].Seg = NULL;
                     Z[j*nf + i].SumOfSegs = 0.0;
											  
                     nEl = 0;
											  
                     x2 = Cosf*dc[j] - Sinfa;
                     y2 = Sinf*dc[j] + Cosfa;

                     if (x1 != x2)
                     {
                        if (y1 != y2)
                        {
                           m = (y2 - y1) / (x2 - x1);
                           b = y1 - m*x1;

                           for (k = 0; k < nyPlusOne; k++)
                           {
                              xhrz = (xl[k] - b) / m;
                              yhrz = xl[k];
                              if ((xhrz >= xLoLimit) &&
                                  (xhrz <= xUpLimit) &&
                                  (yhrz >= yLoLimit) &&
                                  (yhrz <= yUpLimit))
                              {
                                 V.push_back(make_pair(xhrz, yhrz));
                                 nEl++;
                              }
                           }
                           for (k = 0; k < nxPlusOne; k++)
                           {
                              xvrt = yl[k];
                              yvrt = m*yl[k] + b;
                              if ((xvrt >= xLoLimit) &&
                                  (xvrt <= xUpLimit) &&
                                  (yvrt >= yLoLimit) &&
                                  (yvrt <= yUpLimit))
                              {
                                 V.push_back(make_pair(xvrt, yvrt));
                                 nEl++;
                              }
                           }

                           sort(V.begin(), V.end());
                           for (k = 0; k < nEl; k++)
                           {
                              xV[k] = V[k].first;
                              yV[k] = V[k].second;
                           }

                           if (nEl >= 2)
                           {
                              Z[j*nf+i].Count=nEl-1;
                              Z[j*nf+i].Lin=(int*   )malloc(Z[j*nf+i].Count*sizeof(int   ));
                              Z[j*nf+i].Col=(int*   )malloc(Z[j*nf+i].Count*sizeof(int   ));
                              Z[j*nf+i].Seg=(double*)malloc(Z[j*nf+i].Count*sizeof(double));
                              Z[j*nf+i].ind=(int*   )malloc(Z[j*nf+i].Count*sizeof(int   ));

                              for (k = 1; k < nEl; k++)
                              {
                                 xcm = (xV[k - 1] + xV[k]) / 2.0;
                                 ycm = (yV[k - 1] + yV[k]) / 2.0;
                                 Col = (int)floor(xcm + Var1);
                                 if (Col > nyMinusOne)
                                    Col = Col - 1;
                                 Lin = (int)floor(Var2 - ycm);
                                 if (Lin > nxMinusOne)
                                    Lin = Lin - 1;
                                 seg=sqrt(pow(xV[k]-xV[k-1],2)+pow(yV[k]-yV[k-1],2));
                                 S[j*nf + i] = S[j*nf + i] + seg*Mu1[Lin*ny + Col];

                                 Z[j*nf + i].Lin[k - 1] = Lin;
                                 Z[j*nf + i].Col[k - 1] = Col;
                                 Z[j*nf + i].ind[k - 1] = Lin*ny + Col;
                                 Z[j*nf + i].Seg[k - 1] = seg;
                                 Z[j*nf + i].SumOfSegs += seg;

                                 mO1[Lin*ny + Col] = mO1[Lin*ny + Col] + 1.0;
                              }
                           }
                        }
                        else
                        {
                           if ((y1 >= yLoLimit) && (y1 <= yUpLimit))
                           {
                              Z[j*nf+i].Count=ny;
                              Z[j*nf+i].Lin=(int*   )malloc(Z[j*nf+i].Count*sizeof(int   ));
                              Z[j*nf+i].Col=(int*   )malloc(Z[j*nf+i].Count*sizeof(int   ));
                              Z[j*nf+i].Seg=(double*)malloc(Z[j*nf+i].Count*sizeof(double));
                              Z[j*nf+i].ind=(int*   )malloc(Z[j*nf+i].Count*sizeof(int   ));

                              ycm = y1;
                              Lin = (int)floor(Var2 - ycm);
                              if (Lin > nxMinusOne)
                                 Lin = Lin - 1;
                              for (k = 1; k <= ny; k++)
                              {
                                 xcm = (xl[k - 1] + xl[k]) / 2.0;
                                 Col = (int)floor(xcm + Var1);
                                 if (Col > nyMinusOne)
                                    Col = Col - 1;
                                 S[j*nf + i] = S[j*nf + i] + Mu1[Lin*ny + Col];

                                 Z[j*nf + i].Lin[k - 1] = Lin;
                                 Z[j*nf + i].Col[k - 1] = Col;
                                 Z[j*nf + i].ind[k - 1] = Lin*ny + Col;
                                 Z[j*nf + i].Seg[k - 1] = 1.0;
                                 Z[j*nf + i].SumOfSegs += 1.0;

                                 mO1[Lin*ny + Col] = mO1[Lin*ny + Col] + 1.0;
                              }
                           }
                        }
                     }
                     else
                     {
                        if ((x1 >= xLoLimit) && (x1 <= xUpLimit))
                        {
                           Z[j*nf+i].Count=nx;
                           Z[j*nf+i].Lin=(int*   )malloc(Z[j*nf+i].Count*sizeof(int   ));
                           Z[j*nf+i].Col=(int*   )malloc(Z[j*nf+i].Count*sizeof(int   ));
                           Z[j*nf+i].Seg=(double*)malloc(Z[j*nf+i].Count*sizeof(double));
                           Z[j*nf+i].ind=(int*   )malloc(Z[j*nf+i].Count*sizeof(int   ));

                           xcm = x1;
                           Col = (int)floor(xcm + Var1);
                           if (Col > nyMinusOne)
                              Col = Col - 1;
                           for (k = 1; k <= nx; k++)
                           {
                              ycm = (yl[k - 1] + yl[k]) / 2.0;
                              Lin = (int)floor(Var2 - ycm);
                              if (Lin > nxMinusOne)
                                 Lin = Lin - 1;
                              S[j*nf + i] = S[j*nf + i] + Mu1[Lin*ny + Col];

                              Z[j*nf + i].Lin[k - 1] = Lin;
                              Z[j*nf + i].Col[k - 1] = Col;
                              Z[j*nf + i].ind[k - 1] = Lin*ny + Col;
                              Z[j*nf + i].Seg[k - 1] = 1.0;
                              Z[j*nf + i].SumOfSegs += 1.0;

                              mO1[Lin*ny + Col] = mO1[Lin*ny + Col] + 1.0;
                           }
                        }
                     }
                     if (Z[j*nf + i].Count > 0)
                     for (k = 0; k < Z[j*nf + i].Count; k++)
                        RI1[Z[j*nf+i].Lin[k]*ny+Z[j*nf+i].Col[k]]+=Z[j*nf+i].Seg[k]*
                        (S[j*nf+i]/Z[j*nf+i].SumOfSegs);

                     V.clear();
                  }
               }

               start2 = clock();

               NI = 150;

               InitializeSynchronizationBarrier(&barrier, nThreads, -1);

               HANDLE  hThreadArray[nThreads];
               DWORD   dwThreadIdArray[nThreads];

               for (i = 0; i < nThreads; i++)
               {
                  hThreadArray[i] = CreateThread(
                     NULL,
                     0,
                     MyThreadFunction,
                     IntToPtr(i),
                     0,
                     &dwThreadIdArray[i]);

                  if (hThreadArray[i] == NULL)
                     ExitProcess(3);
               }

               WaitForMultipleObjects(nThreads, hThreadArray, TRUE, INFINITE);

               for (i = 0; i < nThreads; i++)
                  CloseHandle(hThreadArray[i]);

               DeleteSynchronizationBarrier(&barrier);


               duration2 = (double)(clock() - start2) / CLOCKS_PER_SEC;
               duration1 = (double)( start2 - start1) / CLOCKS_PER_SEC;




               min1 = RI1[0];
               for (i = 0; i < nx; i++)
               for (j = 0; j < ny; j++)
               if (RI1[i*ny + j] < min1)
                  min1 = RI1[i*ny + j];

               max1 = RI1[0];
               for (i = 0; i < nx; i++)
               for (j = 0; j < ny; j++)
               if (RI1[i*ny + j] > max1)
                  max1 = RI1[i*ny + j];




               unsigned char c;

               if (cs1)
                  free(cs1);
               cs1 = (unsigned char*)malloc(3 * nx * ny + 2 * nx);

               for (i = 0; i < nx; i++)
               {
                  for (j = 0; j < ny; j++)
                  {
                     c = (unsigned char)(Mu1[i*ny + j]*255.0);
                     cs1[i * 3 * ny + i * 2 + 3 * j] = c;
                     cs1[i * 3 * ny + i * 2 + 3 * j + 1] = c;
                     cs1[i * 3 * ny + i * 2 + 3 * j + 2] = c;
                  }
               }

               if (pixels)
                  free(pixels);
               pixels = (unsigned char*)malloc(3 * nx * ny + 2 * nx);

               for (i = 0; i < nx; i++)
               {
                  for (j = 0; j < ny; j++)
                  {
                     c = (unsigned char)((RI1[i*ny + j]/max1)*255.0);

                     pixels[i * 3 * ny + i * 2 + 3 * j] = c;
                     pixels[i * 3 * ny + i * 2 + 3 * j + 1] = c;
                     pixels[i * 3 * ny + i * 2 + 3 * j + 2] = c;
                  }
               }

               if (Mu1) free(Mu1);
               if (RI1) free(RI1);
               if (RIG) free(RIG);
               if (mO1) free(mO1);

               if (Z)
               for (i = 0; i < np; i++)
               for (j = 0; j < nf; j++)
               {
                  free(Z[i*nf + j].Lin);
                  free(Z[i*nf + j].Col);
                  free(Z[i*nf + j].ind);
                  free(Z[i*nf + j].Seg);
               }
               if (Z)   free(Z);

               if (dc)  free(dc);
               if (xl)  free(xl);
               if (yl)  free(yl);
               if (S)   free(S);
               if (xV)  free(xV);
               if (yV)  free(yV);

               SetCursor(LoadCursor(NULL, IDC_WAIT));
               InvalidateRect(hWnd, NULL, TRUE);
               Status = 1;
               MSG msg1;
               msg1.hwnd = hWnd;
               msg1.message = WM_PAINT;
               DispatchMessage(&msg1);
            }
           break;
         case IDM_ABOUT:
            DialogBox(hInst, MAKEINTRESOURCE(IDD_ABOUTBOX), hWnd, About);
            break;
         case IDM_EXIT:
            DestroyWindow(hWnd);
            break;
         default:
            return DefWindowProc(hWnd, message, wParam, lParam);
         }
         break;
      case WM_PAINT:
         hdc = BeginPaint(hWnd, &ps);
         RECT pRect;

         switch (Status)
         {
            case 0:
               break;
            case 1:
               GetClientRect(hWnd, &pRect);

               bmih.biSize = sizeof(BITMAPINFOHEADER);
               bmih.biWidth = ny;
               bmih.biHeight = -nx;
               bmih.biPlanes = 1;
               bmih.biBitCount = 24;
               bmih.biCompression = BI_RGB;
               bmih.biSizeImage = 0;
               bmih.biXPelsPerMeter = 10;
               bmih.biYPelsPerMeter = 10;
               bmih.biClrUsed = 0;
               bmih.biClrImportant = 0;


               ZeroMemory(&dbmi, sizeof(dbmi));
               dbmi.bmiHeader = bmih;
               dbmi.bmiColors->rgbBlue = 0;
               dbmi.bmiColors->rgbGreen = 0;
               dbmi.bmiColors->rgbRed = 0;
               dbmi.bmiColors->rgbReserved = 0;
               bits = (void*)&(cs1[0]);


               hbmp = CreateDIBSection(hdc, &dbmi, DIB_RGB_COLORS, &bits, NULL, 0);

               if (hbmp == NULL)
                  MessageBox(
                     hWnd,
                     (LPCWSTR)L"Couldn't create bitmap!",
                     (LPCWSTR)L"Error!",
                     MB_OK | MB_ICONEXCLAMATION);
               memcpy(bits, cs1, 3 * nx * ny + 2 * nx);



               hdcMem = CreateCompatibleDC(hdc);
               hbmOld = SelectObject(hdcMem, hbmp);
               GetObject(hbmp, sizeof(bmp), &bmp);
               BitBlt(hdc,pRect.right/2-ny,0,bmp.bmWidth,bmp.bmHeight,hdcMem,0,0,SRCCOPY);
               SelectObject(hdcMem, hbmOld);
               DeleteDC(hdcMem);
               // end displaying Cross-Section


               bmih.biSize = sizeof(BITMAPINFOHEADER);
               bmih.biWidth = ny;
               bmih.biHeight = -nx;
               bmih.biPlanes = 1;
               bmih.biBitCount = 24;
               bmih.biCompression = BI_RGB;
               bmih.biSizeImage = 0;
               bmih.biXPelsPerMeter = 10;
               bmih.biYPelsPerMeter = 10;
               bmih.biClrUsed = 0;
               bmih.biClrImportant = 0;


               ZeroMemory(&dbmi, sizeof(dbmi));
               dbmi.bmiHeader = bmih;
               dbmi.bmiColors->rgbBlue = 0;
               dbmi.bmiColors->rgbGreen = 0;
               dbmi.bmiColors->rgbRed = 0;
               dbmi.bmiColors->rgbReserved = 0;
               bits = (void*)&(pixels[0]);


               hbmp = CreateDIBSection(hdc, &dbmi, DIB_RGB_COLORS, &bits, NULL, 0);

               if (hbmp == NULL)
                  MessageBox(
                     hWnd,
                     (LPCWSTR)L"Couldn't create bitmap!",
                     (LPCWSTR)L"Error!",
                     MB_OK | MB_ICONEXCLAMATION);
               memcpy(bits, pixels, 3 * nx * ny + 2 * nx);



               hdcMem = CreateCompatibleDC(hdc);
               hbmOld = SelectObject(hdcMem, hbmp);
               GetObject(hbmp, sizeof(bmp), &bmp);
               BitBlt(hdc,pRect.right/2,0,bmp.bmWidth,bmp.bmHeight,hdcMem,0,0,SRCCOPY);
               SelectObject(hdcMem, hbmOld);
               DeleteDC(hdcMem);
               // end displaying reconstruction

               SetTextColor(hdc, RGB(255, 0, 0));

               RECT rep, r1, r2, r3, r4;
               rep.left = 0;
               rep.top = nx + 10;
               rep.right = pRect.right;
               rep.bottom = nx + 40;
               r1.left = 0;
               r1.top = nx + 40;
               r1.right = pRect.right / 2;
               r1.bottom = nx + 70;
               r2.left = 0;
               r2.top = nx + 70;
               r2.right = pRect.right / 2;
               r2.bottom = nx + 100;
               r3.left = 0;
               r3.top = nx + 100;
               r3.right = pRect.right / 2;
               r3.bottom = nx + 130;
               r4.left = 0;
               r4.top = nx + 130;
               r4.right = pRect.right / 2;
               r4.bottom = nx + 160;

               WCHAR buffer1[50];
               int len;

               len = swprintf(buffer1, 50, L"Report");
               DrawText(hdc, (LPTSTR)buffer1, len, &rep, DT_CENTER);

               len = swprintf(buffer1, 50, L"Time 1:");
               DrawText(hdc, (LPTSTR)buffer1, len, &r1, DT_RIGHT);

               len = swprintf(buffer1, 50, L"Time 2:");
               DrawText(hdc, (LPTSTR)buffer1, len, &r2, DT_RIGHT);

               len = swprintf(buffer1, 50, L"   V1:");
               DrawText(hdc, (LPTSTR)buffer1, len, &r3, DT_RIGHT);

               len = swprintf(buffer1, 50, L"   V2:");
               DrawText(hdc, (LPTSTR)buffer1, len, &r4, DT_RIGHT);

               RECT r5, r6, r7, r8;
               r5.left = pRect.right / 2;
               r5.top = nx + 40;
               r5.right = pRect.right / 2 + 100;
               r5.bottom = nx + 70;
               r6.left = pRect.right / 2;
               r6.top = nx + 70;
               r6.right = pRect.right / 2 + 100;
               r6.bottom = nx + 100;
               r7.left = pRect.right / 2;
               r7.top = nx + 100;
               r7.right = pRect.right / 2 + 100;
               r7.bottom = nx + 130;
               r8.left = pRect.right / 2;
               r8.top = nx + 130;
               r8.right = pRect.right / 2 + 100;
               r8.bottom = nx + 160;

               len = swprintf(buffer1, 50, L"%12.5f", duration1);
               DrawText(hdc, (LPTSTR)buffer1, len, &r5, DT_RIGHT);

               len = swprintf(buffer1, 50, L"%12.5f", duration2);
               DrawText(hdc, (LPTSTR)buffer1, len, &r6, DT_RIGHT);

               len = swprintf(buffer1, 50, L"%12.5f", min1);
               DrawText(hdc, (LPTSTR)buffer1, len, &r7, DT_RIGHT);

               len = swprintf(buffer1, 50, L"%12.5f", max1);
               DrawText(hdc, (LPTSTR)buffer1, len, &r8, DT_RIGHT);

               MoveToEx(hdc, 0, nx + 25, NULL);
               LineTo(hdc, pRect.right, nx + 25);
               MoveToEx(hdc, 0, nx + 55, NULL);
               LineTo(hdc, pRect.right, nx + 55);
               MoveToEx(hdc, 0, nx + 85, NULL);
               LineTo(hdc, pRect.right, nx + 85);
               MoveToEx(hdc, 0, nx + 115, NULL);
               LineTo(hdc, pRect.right, nx + 115);
               MoveToEx(hdc, 0, nx + 145, NULL);
               LineTo(hdc, pRect.right, nx + 145);

               break;
            case 2:
               GetClientRect(hWnd, &pRect);
               DrawText(hdc, (LPTSTR)Msg1, 50, &pRect, DT_CENTER);
               break;
            default:
               break;
         }
         EndPaint(hWnd, &ps);
         break;
      case WM_DESTROY:
         DeleteObject(hbmp);
         PostQuitMessage(0);
         break;
      default:
         return DefWindowProc(hWnd, message, wParam, lParam);
   }
   
   return 0;
}

// Message handler for About box
INT_PTR CALLBACK About(HWND hDlg, UINT message, WPARAM wParam, LPARAM lParam)
{
   UNREFERENCED_PARAMETER(lParam);
   switch (message)
   {
      case WM_INITDIALOG:
         return (INT_PTR)TRUE;
      case WM_COMMAND:
         if (LOWORD(wParam) == IDOK || LOWORD(wParam) == IDCANCEL)
         {
            EndDialog(hDlg, LOWORD(wParam));
            return (INT_PTR)TRUE;
         }
         break;
   }
   return (INT_PTR)FALSE;
}

DWORD WINAPI MyThreadFunction(LPVOID lpParam)
{
   int id = PtrToInt(lpParam);

   int i;
   int j;
   int k;
   int z;

   int	   *pI;
   double *pS;

   int index1;
   int index2;
   int index3;

   double Sit;
   double aux1;

   double *RI2 = (double*)malloc(nx*ny*sizeof(double));
   for (i = 0; i < nx; i++)
   for (j = 0; j < ny; j++)
   {
      RI2[i*ny + j] = 0.0;
   }

   int i1;
   int i2;

   switch (id)
   {
      case 3:
         i1 = 0;
         i2 = 45;
         break;
      case 2:
         i1 = 45;
         i2 = 90;
         break;
      case 1:
         i1 = 90;
         i2 = 135;
         break;
      case 0:
         i1 = 135;
         i2 = 180;
         break;
      default:
         break;
   }

   for (z = NI; z != 0; z--)
   {
      for (i = i1; i < i2; i++)
      {
         for (j = 0; j < np; j++)
         {
            if (Z[j*nf + i].Count > 0)
            {
               index1 = j*nf + i;
               index2 = Z[index1].Count;

               pI = Z[index1].ind;
               pS = Z[index1].Seg;

               Sit = 0.0;
               for (k = 0; k < index2; k++)
               {
                  Sit += (*pS) * RI1[*pI];
                  pI++;
                  pS++;
               }

               if (Sit > 0.0)
               {
                  aux1 = S[index1] / Sit;

                  pI = Z[index1].ind;

                  for (k = 0; k < index2; k++)
                  {
                     index3 = (*pI);
                     RI2[index3] += RI1[index3] * aux1;

                     pI++;
                  }
               }
               else
               {
                  if (S[index1] > 0.0)
                  {
                     MessageBox(NULL, (LPCWSTR)L"!!", (LPCWSTR)L"!", MB_OK);
                     pI = Z[index1].ind;
                     pS = Z[index1].Seg;
                     for (k = 0; k < index2; k++)
                     {
                        RI2[(*pI)] += (*pS)*(S[index1] / Z[index1].SumOfSegs);
                        pI++;
                        pS++;
                     }
                  }
               }
            }
         }
      }

		
      if (id == 3)
      {
         for (i = 0; i < nx; i++)
         for (j = 0; j < ny; j++)
         {
            RIG[i*ny + j] += RI2[i*ny + j];
         }
      }

      EnterSynchronizationBarrier(&barrier, 0);

      if (id == 2)
      {
         for (i = 0; i < nx; i++)
         for (j = 0; j < ny; j++)
         {
            RIG[i*ny + j] += RI2[i*ny + j];
         }
      }

      EnterSynchronizationBarrier(&barrier, 0);

      if (id == 1)
      {
         for (i = 0; i < nx; i++)
         for (j = 0; j < ny; j++)
         {
            RIG[i*ny + j] += RI2[i*ny + j];
         }
      }

      EnterSynchronizationBarrier(&barrier, 0);

      if (id == 0)
      {
         for (i = 0; i < nx; i++)
         for (j = 0; j < ny; j++)
         {
            RIG[i*ny + j]+= RI2[i*ny + j];
            RI1[i*ny + j] = RIG[i*ny + j] / mO1[i*ny + j];
            RIG[i*ny + j] = 0.0;
         }
      }

      EnterSynchronizationBarrier(&barrier, 0);

      for (i = 0; i < nx; i++)
      for (j = 0; j < ny; j++)
      {
         RI2[i*ny + j] = 0.0;
      }
   }

   return 0;
}
\end{lstlisting}

\end{document}